\documentclass{article}
\usepackage[utf8]{inputenc}
\usepackage{graphicx}
\usepackage{epsfig}
\pdfoutput=1
\usepackage{amsmath}
\usepackage{amsthm}
\usepackage{amssymb}
\usepackage{xcolor}
\usepackage{cite}

\title{The nonlinear Klein-Gordon equation on metric graphs: Modeling reflectionless transmission of the kink soliton}
\author{
Asadov Q.U.*, Sabirov K.K.**, Aripov M.*** \\
\textit{*, ** Tashkent university of information technologies
named after Al-Khwarizmi} \\
\textit{Dept. Algoritmization and mathematical modelling} \\
\textit{*** National University of Uzbekistan named after Mirzo Ulugbek} \\
\textit{Dept. Applied mathematics and computer analysis} \\
\textit{Tashkent, Uzbekistan}
}

\date{\today}

\begin{document}

\maketitle
\begin{abstract}
In this paper we consider the nonlinear Klein-Gordon equation on the metric star graph with tree semi-infinite bonds. At the branched point we put two types of vertex boundary conditions: the weight continuity and the condition for derivatives of wave functions as the generalized Kirchhoff rule. We solve this equation satisfying vertex boundary conditions and the energy, momentum conservation laws. We also show reflectionsless propagations of the kink soliton solution, plot the reflection coefficient and extend to other topologies.
\end{abstract}
\section{Introduction}

Nonlinear dynamics of solitary waves, that occur in many scientific areas, can be described by nonlinear partial differential equations \cite{Ablowitz1}, \cite{Ablowitz2}. The nonlinear Klein-Gordon equation is one of the important classes of partial differential equations and is used to model many different phenomena, including the propagation of dislocations in crystals and the
behavior of elementary particles \cite{Greiner}. The exact and numerical soliton solutions of Klein-Gordon equation are given in the pioneer work \cite{Ablowitz3}, in that numerically solved with the time and coordinate steps $k=h=0.05$ and found the descritization scheme using Taylor series. The energy and momentum conservating schemes  used to integrate the nonlinear Klein-Gordon equation in \cite{Jim} and invariant-conserving finite difference algorithms are given in \cite{Los}.

The nonlinear partial differential equations on the branched structures are attached the most interest in the last decade years. Such attention was caused by the possibility for obtaining soliton solution of nonlinear partial differential equations such as nonlinear Schr\"odinger, Dirac equations and its numerous applications in different branches of physics  \cite{Sobirov1,Adami,Sobirov2,Sabirov1,Sabirov2}. The branched structures can be modeled metric graphs. Metric graphs are consist of two sets: the set of points (vertices) and the set of intervals that connect vertices. Topology of graphs can be given so called adjacency matrices \cite{Kottos,Gnutzmann}:
$$
C_{ij}=\left\{\begin{array}{ll}1,\,i,j\text{ are connected,}\\0,\,\text{otherwise,}\end{array}\right.i,j=1,2,...,V
$$

In this paper we focus on one of the exact solutions and transmission
of the kink soliton of the nonlinear Klein-Gordon equation through the vertices of the networks. The present work is organized as follows: in the section 2  the nonlinear Klein-Gordon equation on the metric star graph with derivations the vertex boundary conditions is obtained; in the section 3 the kink soliton solution of the formulated problem with the sum rule for the reflectionless transmission is analytically given, the momentum conservation law is shown constant and plotted the energy conservation law with the plotting reflection coefficient as the ratio of momentums; the section 4 is devoted for the extensions to other topologies of metric graphs such as the tree and loop graphs; in the last section the conclusions are given.

\section{Formulation of the problem}
We consider a star graph with three bonds $e_j$, for which a coordinate $x_j$ is assigned. Choosing the origin of coordinates at the vertex, 0, for bond $e_1$ we put $x_1\in(-\infty,0]$ and for $e_{2,3}$ we fix $x_{2,3}\in[0,+\infty)$. In what follows, we use the shorthand notation $q_{j}(x)$ for $q_j(x_j)$ where $x$ is the coordinate on the bond $j$ to which the component $q_j$ refers. Klein-Gordon equation on the each bond $e_j$ of the star graph is written as
\begin{equation}
\partial_{tt}^2 q_j - \partial_{xx}^2 q_j - q_j + b_{j} q_{j}^3 =
0.\label{kgeq1}
\end{equation}

\begin{figure}[ht!]
\centering \includegraphics[scale = 0.5]{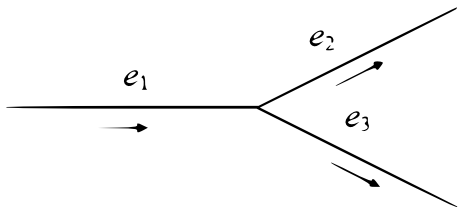}
\caption{The metric star graph} \label{pic1}
\end{figure}
Now we define the vertex boundary conditions at the branched point of the star graph, therefore we derive these boundary conditions from conservation laws. One of the conservation laws is energy. The energy conservation law is defined as
\begin{equation}
E = \sum_{j=1}^3 E_j,\label{conl1}
\end{equation}
where
\begin{equation}
E_j= \int _{e_j} \left[ \frac{1}{2} (\partial_t
q_j)^2 + \frac{1}{2} (\partial_x q_j)^2 - \frac{1}{2} q_j^2 +
\frac{b_j}{4} q_j^4 \right] dx.\label{energy1}
\end{equation}
From $\dot{E}=0$ we can get the following nonlinear boundary condition as
\begin{equation}
\partial_x q_1 \partial_t q_1 |_{x=0} = \partial_x q_2 \partial_t q_2 |_{x=0} + \partial_x q_3 \partial_t q_3|_{x=0}.\label{nbc1}
\end{equation}
We need two types of boundary conditions to find a solution of (\ref{kgeq1}) and to fulfil the nonlinear vertex boundary condition (\ref{nbc1}). Therefore the first type of vertex boundary conditions is the following weight continuity
\begin{equation}
 \alpha_1 q_1 |_{x=0}= \alpha_2 q_2 |_{x=0} = \alpha_3 q_3 |_{x=0},\label{wc1}
 \end{equation}
 the second type of vertex boundary conditions is given derivatives of wave functions at the branched point as Kirchhoff rule
\begin{equation}
 \frac{1}{\alpha_1} \partial_x q_1 |_{x=0} = \frac{1}{\alpha_2} \partial_x q_2 |_{x=0} + \frac{1}{\alpha_3} \partial_x q_3 |_{x=0}.\label{kr1}
\end{equation}

\section{The kink soliton solution of Klain-Gordon equation on the star graph with three edges}

The kink (antikink) soliton solution of Klein-Gordon equation (\ref{kgeq1}) on the each bond $e_j$ of the metric star graph is the following
\begin{equation}
 q_j (x,t) = \mp \frac{1}{\sqrt{b_j}} \, {\rm tanh} \left( \frac{x - l - \upsilon t}
{\sqrt{2(1 - \upsilon^2) }} \right),\label{sol1}
\end{equation}
where $l$ is the initial center of mass of soliton (the kink and antikink soliton solutions are with the signs $-$ and $+$, respectively). Fulfilling the vertex boundary conditions (\ref{wc1})-(\ref{kr1}) we can get
the following constrains
\begin{eqnarray}
\frac {\alpha_1}{\sqrt{b_1}} = \frac {\alpha_2}{\sqrt{b_2}} =
\frac{\alpha_3}{\sqrt{b_3}},\label{const1}\\
\frac {1}{\alpha_1 \sqrt{b_1}} = \frac {1}{\alpha_2 \sqrt{b_2}} +
\frac{1}{\alpha_3 \sqrt{b_3}}.\label{const2}
\end{eqnarray}
From Eq.s(\ref{const1}) and (\ref{const2}) we obtain the following
sum rule for nonlinearities
\begin{equation}
\frac {1}{b_1} = \frac {1}{b_2} + \frac {1}{b_3}.\label{sr1}
\end{equation}

\begin{figure}[h!]
\begin{minipage} [h!]{0.5\linewidth}
\centering \includegraphics[scale=0.45]{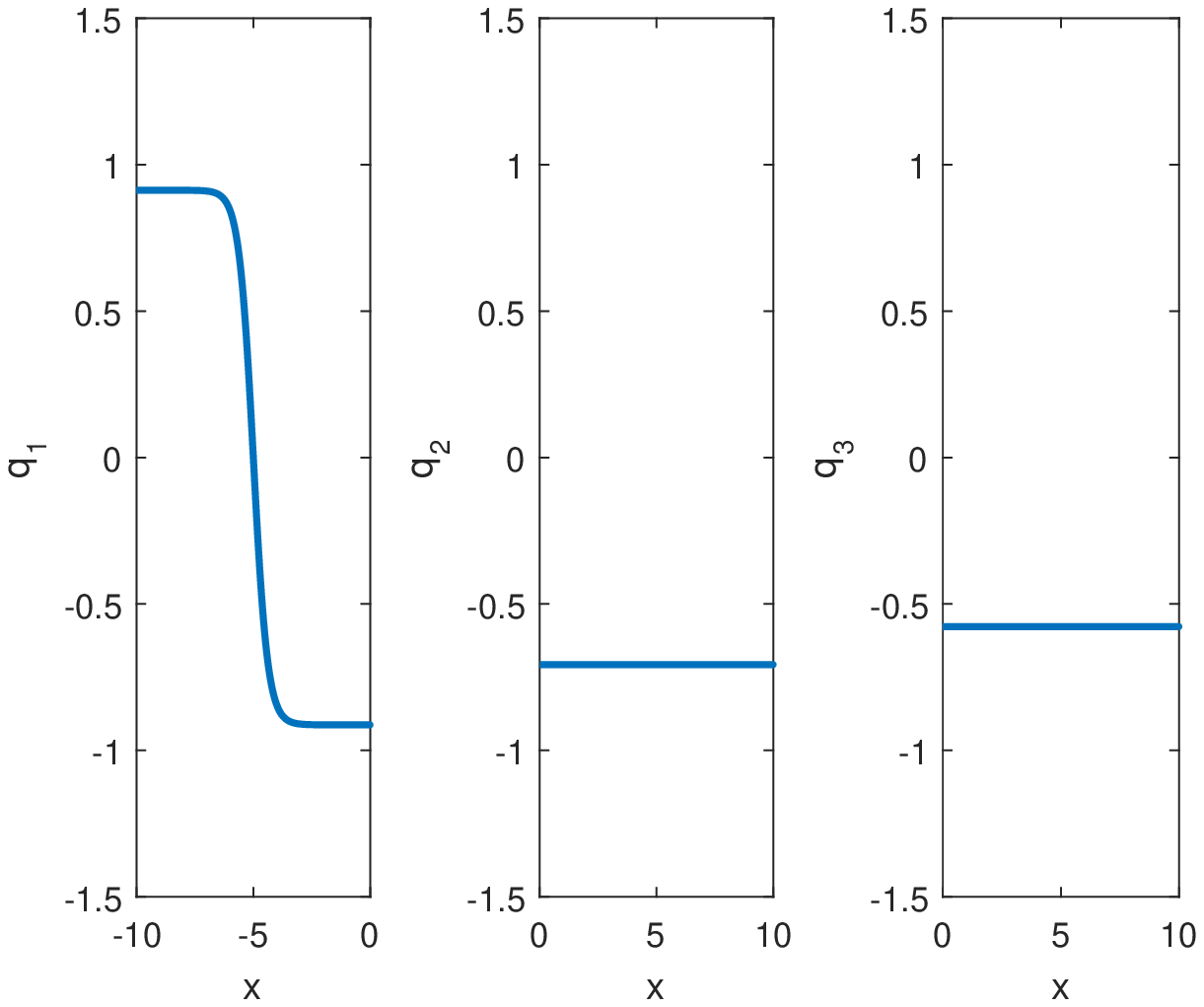} \\ a)
\end{minipage}
\begin{minipage}[h!]{0.5\linewidth}
\centering \includegraphics[scale=0.45]{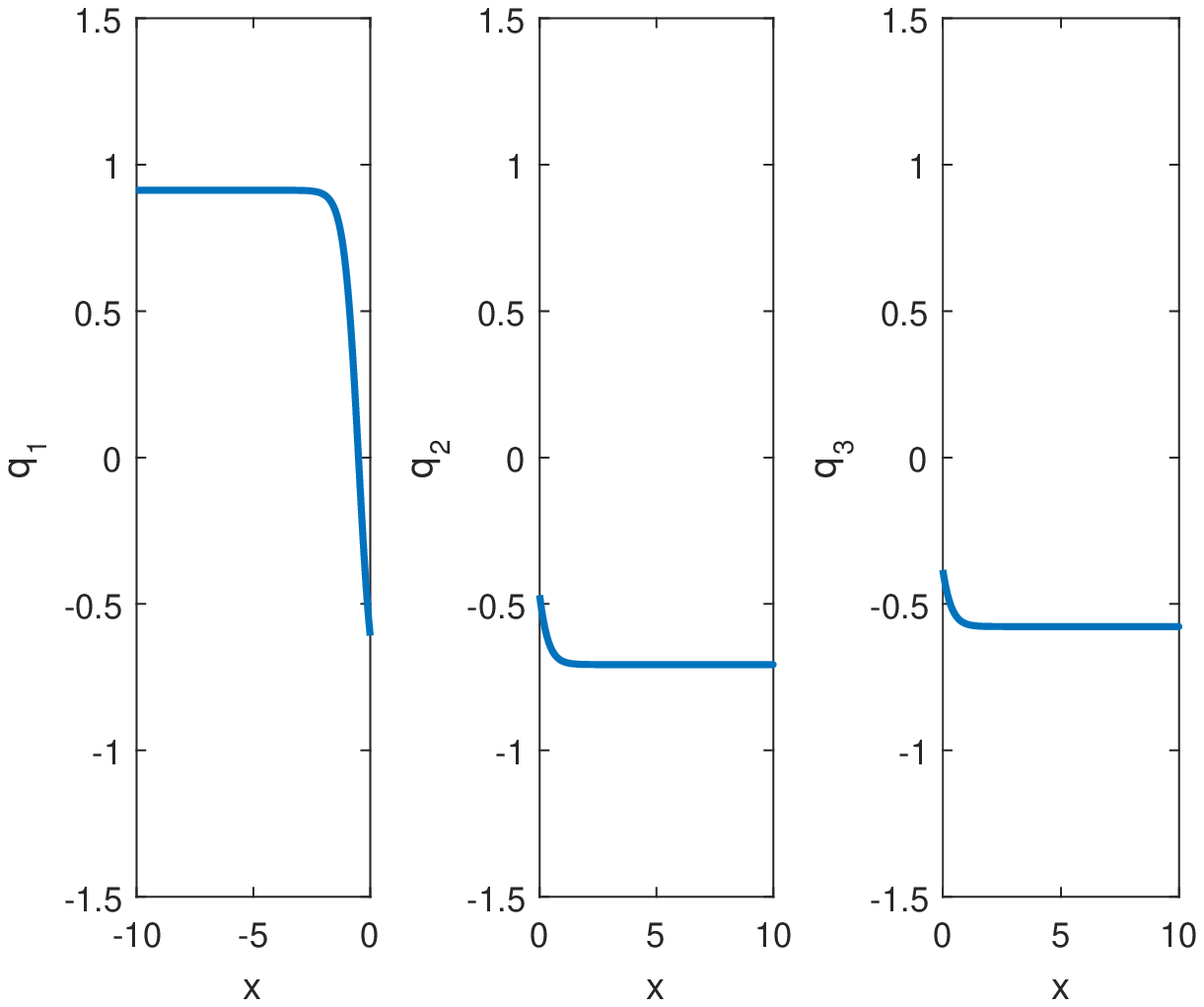} \\ b)
\end{minipage}
\begin{minipage}[h!]{\linewidth}
\centering \includegraphics[scale=0.5]{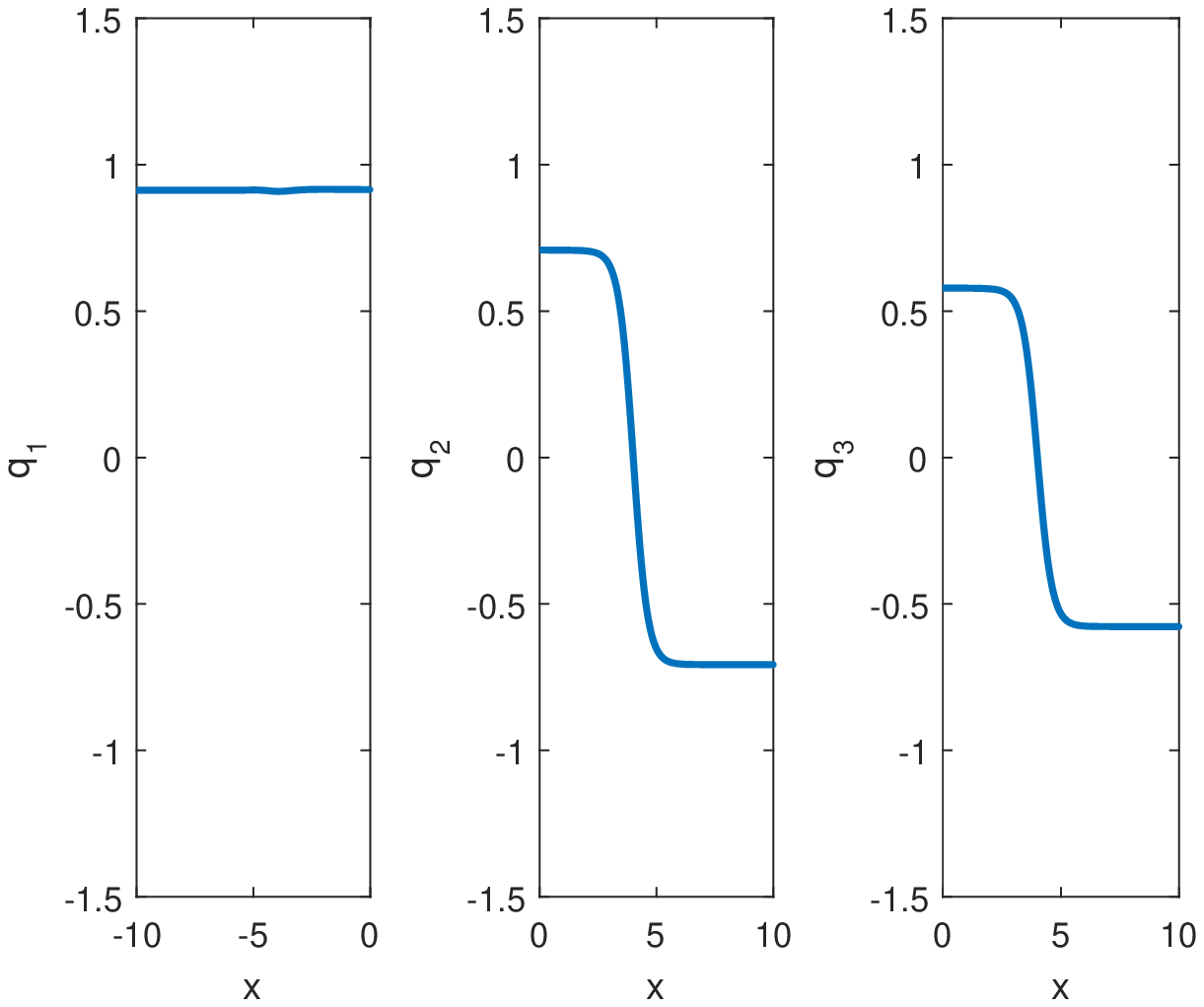} \\ c)
\end{minipage}

\caption{Reflectionless Transmission: a) t=0; b) t=5; c) t=10.} \label{pic2}
\end{figure}

Using the kink (antikink) soliton solution of Klein-Gordon equation (\ref{kgeq1}) with the sum rule for the nonlinearities (\ref{sr1}) we can also show that another conservation law, i.e. the momentum is conserved:
$$
P = \sum _{j=1}^3 \int_{e_j} \partial _t q_j \partial _x q_j dx = \sum _{j=1}^3 \frac{1}{2 b_j} \cdot \frac { \upsilon } { 1 -
\upsilon ^2 } \int_{e_j} \frac{dx}{cosh ^4 \left( \frac{ x - l
-\upsilon t} {\sqrt{2(1-\upsilon^2)}} \right)} =
$$
$$
= \frac { \upsilon } { 2(1 - \upsilon ^2) }\left[ \frac{1}{b_1}
\int_{-\infty} ^{0} \frac{dx}{cosh ^4 \left( \frac{ x - l
-\upsilon t} {\sqrt{2(1-\upsilon^2)}} \right)} +
\left(\frac{1}{b_2} + \frac{1}{b_3} \right) \int_0 ^{+\infty}
\frac{dx}{cosh ^4 \left( \frac{ x - l -\upsilon t}
{\sqrt{2(1-\upsilon^2)}} \right)}\right] =
$$
$$
= \frac { \upsilon } {2 (1 - \upsilon ^2) b_1 } \int_{-\infty}
^{+\infty} \frac{dx}{cosh ^4 \left( \frac{ x - l -\upsilon t}
{\sqrt{2(1-\upsilon^2)}} \right)} = \frac{4 \upsilon }{ 3 b_1 \sqrt{2(1-\upsilon^2)} },
$$
from the last expression the momentum is constant.

\begin{figure}[h!]
\centering \includegraphics[width=8.6cm]{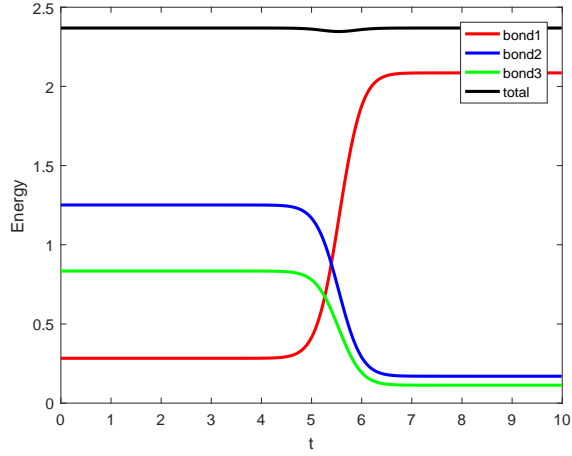}
\caption{The energies on the each bonds and the total energy} \label{pic3}
\end{figure}

In the figure \ref{pic2} the propagation of the kink soliton solution from the first bond to the second and third bonds is plotted and the refletionless transmission is shown at $t=0,\,t=5$ and $t=10$. In the figure \ref{pic3} it is plotted the energies on the each bonds and the total energy, from this figure it is clear seen that the total energy is constant. The reflection coefficient $R(b_3,t=10)=\frac{P_1(t=10)}{P_1(t=10)+P_2(t=10)+P_3(t=10)}$ is plotted as a function of $b_3$  for fixed $b_j,\,j=1,2,\,\alpha_1,\,\alpha_2,\,\alpha_3=\frac{1}{\sqrt{b3}\left(\frac{1}{\alpha_3\sqrt{b3}}-\frac{1}{\alpha_2\sqrt{b2}}\right)}$. This systematic analysis clearly illustrates the necessity for reflection to be absent at $b_3=3$.

\begin{figure}[h!]
\centering \includegraphics[width=8.6cm]{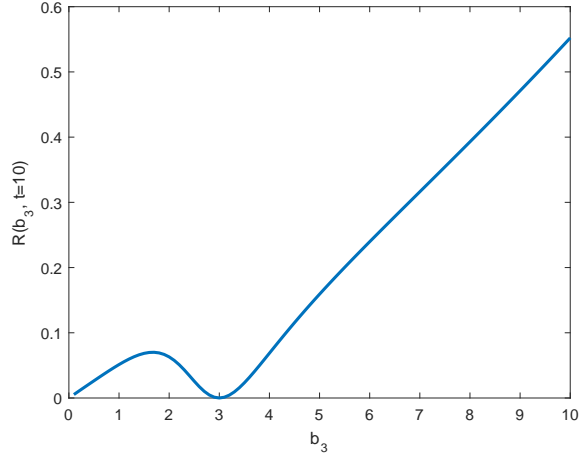}
\caption{The reflection coefficient} \label{pic4}
\end{figure}

\section{Extensions to other topologies of metric graphs}

\begin{figure}[h!]
\centering \includegraphics[scale=0.5]{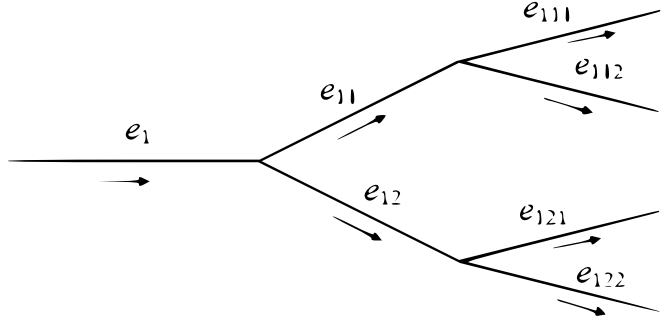} \caption{The
metric tree graph} \label{pic5}
\end{figure}

The above approach of the nonlinear Klein-Gordon equation on the metric star graph can be extended to the tree graph presented in figure \ref{pic5}. This tree graph consists of three subgraphs $e_1 \sim(-\infty;0],\, e_{1i}\sim[0;L_i]$ and $e_{1ij}\sim[0;+\infty)$, where $i,j=1,2$. On the each bond we consider the nonlinear Klein-Gordon equation as given by
\begin{equation}
\partial_{tt}^2 q_e - \partial_{xx}^2 q_e - q_e - b_e q_e^3=0, \label{kgeq2}
\end{equation}
with the following boundary conditions at the vertices
\begin{eqnarray}
\left\{%
\begin{array}{ll}
 \alpha_1 q_1 |_{x=0}= \alpha_{11} q_{11} |_{x=0} = \alpha_{12} q_{12} |_{x=0} ; \\
 \frac{1}{\alpha_1} \partial_x q_1 |_{x=0} = \frac{1}{\alpha_{11}} \partial_x q_{11} |_{x=0} + \frac{1}{\alpha_{12}} \partial_x q_{12} |_{x=0}. \\
 \alpha_{11} q_{11} |_{x=L_1} = \alpha_{111} q_{111} |_{x=0} = \alpha_{112} q_{112} |_{x=0} ; \\
 \frac{1}{\alpha_{11}} \partial_x q_{11} |_{x=L_1} = \frac{1}{\alpha_{111}} \partial_x q_{111} |_{x=0} + \frac{1}{\alpha_{112}} \partial_x q_{112} |_{x=0}. \\
 \alpha_{12} q_{12} |_{x=L_2}= \alpha_{121} q_{121} |_{x=0} = \alpha_{122} q_{122} |_{x=0} ; \\
 \frac{1}{\alpha_{12}} \partial_x q_{12} |_{x=L_2} = \frac{1}{\alpha_{121}} \partial_x q_{121} |_{x=0} + \frac{1}{\alpha_{122}} \partial_x q_{122} |_{x=0}. \\
\end{array}%
\right.\label{bc1}
\end{eqnarray}
The kink (antikink) soliton solution of equation (\ref{kgeq2}) can be written as
\begin{equation}
q_e (x,t) =\mp \frac{1}{\sqrt{b_e}} \, {\rm tanh} \left( \frac{x - x_{0,e} - \upsilon
t}{\sqrt{2(1 - \upsilon^2) }} \right)\label{sol2}.
\end{equation}
where
\begin{equation}
x_{0,1} = x_{0,11} = x_{0,12} = l,\,x_{0,111} = x_{0,112} = l - L_1,\,x_{0,121} = x_{0,122} = l - L_2.
\end{equation}
$ l -$ initial center of the mass of solution. Satisfying the boundary conditions (\ref{bc1}) we have the following sum rules for nonlinearities
\begin{eqnarray}
\left.\begin{array}{lll}\frac {1}{b_1} = \frac {1}{b_{11}} + \frac {1}{b_{12}},\\
\frac {1}{b_{11}} = \frac {1}{b_{111}} + \frac {1}{b_{112}},\\
\frac {1}{b_{12}} = \frac {1}{b_{121}} + \frac {1}{b_{122}}.
\end{array}\right.\label{sr2}
\end{eqnarray}

\begin{figure}[ht!]
\centering \includegraphics[scale=0.5]{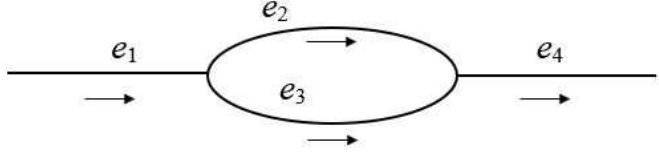}
\caption{The metric loop graph} \label{pic6}
\end{figure}

Now we consider the loop graph plotted in figure \ref{pic6} with subgraphs $e_1\sim(- \infty ; 0] , \, e_{2}, e_3 \sim [0; L] $ and $ e_4 \sim [L; + \infty) $. On the each bond of this loop graph the nonlinear Klein-Gordon equation can be written by equation (\ref{kgeq2}) with the boundary conditions at the vertices

\begin{equation}
\left\{%
\begin{array}{ll}
 \alpha_1 q_1 |_{x=0}= \alpha_{2} q_{2} |_{x=0} = \alpha_{3} q_{3} |_{x=0} ; \\
 \frac{1}{\alpha_1} \partial_x q_1 |_{x=0} = \frac{1}{\alpha_{2}} \partial_x q_{2} |_{x=0} + \frac{1}{\alpha_{3}} \partial_x q_{3} |_{x=0}. \\
 \alpha_{2} q_{2} |_{x=L} = \alpha_{3} q_{3} |_{x=L} = \alpha_{4} q_{4} |_{x=0} ; \\
 \frac{1}{\alpha_{2}} \partial_x q_{2} |_{x=L} + \frac{1}{\alpha_{3}} \partial_x q_{3} |_{x=L} = \frac{1}{\alpha_{4}} \partial_x q_{4} |_{x=0}. \label{bc2}
 \end{array}%
\right.
\end{equation}
The kink (antikink) soliton solution is given by equation (\ref{sol2}), where $x_{0,1} = x_{0,2} = x_{0,3} = l,\,x_{0,4} = l - L$. Satisfying the boundary conditions (\ref{bc2}) we can get the sum rule written as
\begin{eqnarray}
\left.\begin{array}{ll}\frac {1}{b_1} = \frac {1}{b_{2}} + \frac {1}{b_{3}}=\frac {1}{b_{4}}.
\end{array}\right.
\end{eqnarray}
\section{Conclusions}

In this paper we studied the nonlinear Klein-Gordon equation on the simplest metric graphs as the star graph with tree semi-infinite bonds, tree and loop graphs. First of all we derived the nonlinear boundary condition from the energy conservation law. Satisfying this boundary conditions at the vertex (branched point) we obtained the weight continuity and the condition for derivatives of the wave function as Kirchhoff rule. We obtained the soliton solution on the metric star graph and the constrain as inverses of nonlinearities for the reflectionless transmission. Using the soliton solution we got that the total momentum is conserved. We showed the conservation of the total energy and calculated the reflection coefficient as the ratio of the momentum on the first bond to the total momentum in the figures \ref{pic3} and \ref{pic4}, respectively. We also extended obtained results as the formulation problems and the constrains for the nonlinearities to the tree and loop graphs.

\end{document}